\documentclass[%
prl,twocolumn,showpacs
 amsmath,amssymb,
 aps,
]{revtex4-1}

\usepackage{graphicx}
\usepackage{dcolumn}
\usepackage{bm}

\newcommand{\EQ}{\begin{equation}}
\newcommand{\EN}{\end{equation}}
\newcommand{\be}{\begin{equation}}
\newcommand{\ee}{\end{equation}}
\newcommand{\bea}{\begin{eqnarray}}
\newcommand{\eea}{\end{eqnarray}}

\begin{document}


\title{Exact results for quenched bond randomness at criticality}

\author{Gesualdo Delfino}
 \altaffiliation[]{delfino@sissa.it}
\affiliation{SISSA - International School for Advanced Studies, via Bonomea 265, 34136 Trieste, Italy \\
INFN - Istituto Nazionale di Fisica Nucleare, Sezione di Trieste, Italy}


\begin{abstract}
We introduce an exact replica method for the study of critical systems with quenched bond randomness in two dimensions. For the $q$-state Potts model we show that a line of renormalization group fixed points interpolates from weak to strong randomness as $q-2$ grows from small to large values. This theory exhibits a $q$-independent sector, and allows at the same time for a correlation length exponent which keeps the Ising value and continuously varying magnetization exponent and effective central charge. These findings appear to solve long-standing numerical and theoretical puzzles, and to illustrate the peculiarities which may characterize the conformal field theories of random fixed points. 
\end{abstract}

\maketitle




Quenched bond randomness plays an interesting role within the theory of critical phenomena. The Harris criterion \cite{Harris} says that when the specific heat critical exponent $\alpha$ of the pure model is positive weak randomness is relevant in the renormalization group sense and drives the system towards a new (random) fixed point (FP). For weakly relevant randomness, perturbation theory can be used to study these new FPs but, due to its approximate nature, can hardly establish whether they possess distinctive features with respect to those of non-quenched systems. Looking for non-perturbative methods, attention turns to the two-dimensional (2D) case, in which conformal field theory (CFT) has provided an exact and essentially complete characterization of universality classes of critical behavior for pure systems. It is a fact, however, that no CFT for 2D systems with quenched disorder has been identified. On the other hand, conformal invariance for this type of criticality is expected to apply to a larger spectrum of models than in the pure case. Indeed, it has been argued \cite{HB}, and also rigorously shown for a large class of models \cite{AW}, that in 2D bond randomness softens first order phase transitions into continuous ones. 

The 2D $q$-state Potts ferromagnet has played a central role in the study of quenched bond randomness. The model can be continued to real values of $q$ \cite{FK} and in the pure case exhibits a phase transition which becomes first order for $q>4$ \cite{Baxter}. Weak bond randomness, which for the Ising case ($q=2$) is marginally irrelevant \cite{DD} and unable to produce a new FP, becomes relevant for $q>2$. The perturbative analysis for $q\to 2$ yields a line of FPs with varying critical exponents \cite{Ludwig,DPP}. On the other hand, Monte Carlo simulations performed at $q=8$, while confirming the softening of the transition, found exponents consistent with Ising values \cite{CFL}, and a similar conclusion was obtained from simulations at $q=4$ \cite{DW}. A simplified interfacial model \cite{KSSD} then indicated $q$-independence, at least for $q$ sufficiently large, of the interfacial free energy exponent $\mu$ (related to the correlation length exponent $\nu$), with a value numerically consistent with the Ising one. The authors of \cite{KSSD} observed that their analysis involves strong randomness and possibly yields a line of FPs different from that studied perturbatively in \cite{Ludwig,DPP}. This possibility was also suggested in \cite{CJ}, where a numerical transfer matrix study in the range $2\leq q\leq 8$, while finding a very weak $q$-dependence for $\nu$,  established a macroscopic deviation of the magnetization exponent $\beta$ from the Ising value at $q=8$. Meanwhile the $q$-dependence of the effective central charge $c^\prime$ had been found numerically in \cite{Picco}. Following numerical studies \cite{JP}, exact asymptotic values for the exponents at $q=\infty$ have been proposed in \cite{AdI}, in particular the Ising value $\nu=1$. 

In this paper we introduce an {\it exact} replica method for the study of renormalization group FPs of 2D systems with quenched bond randomness. For the $q$-state Potts model we find that all the above mentioned results actually correspond to the {\it same} line of FPs, for which the randomness strength grows from weak to strong as $q-2$ grows from small to large values. Remarkably, this critical line possesses a symmetry-independent sector, and allows at the same time for constant $\nu$ and $q$-dependent $\beta$ and $c^\prime$. The peculiarity of these features makes less surprising that the CFTs of random FPs have not been identified among those solved so far. 

We exploit the scale invariant scattering formalism introduced in \cite{paraf}, where it was illustrated for the cases of the pure $q$-state Potts and $n$-vector models. FPs of the renormalization group for 2D statistical models with short range interactions are identified, directly in the continuum limit, as scale invariant $S$-matrix solutions for the underlying relativistic quantum field theories in (1+1)-dimensional space-time. Relativistic invariance of the quantum theory corresponds to isotropy of the statistical system in the scaling limit. Scale invariance implies massless particles which in 2D are right/left-movers with energy and momentum related as $e=\pm p$. A 2D peculiarity is that conformal invariance, which for local field theories is implied by scale invariance, ensures the presence of an infinite number of integrals of motion forcing the scattering to be completely elastic: the initial and final states contain the same number of particles with the same momenta. The only relativistic invariant in the scattering of a right-mover with a left-mover is the center of mass energy, which is dimensionful; scale invariance and unitarity then imply momentum-independence of the amplitude. As a consequence, the unitarity and crossing symmetry equations \cite{ELOP} take a simple form \cite{paraf,fpu}. 

\begin{figure}
\begin{center}
\includegraphics[width=8cm]{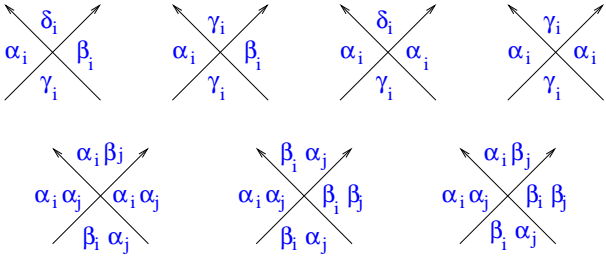}
\caption{Scattering processes corresponding to the amplitudes $S_0$, $S_1$, $S_2$, $S_3$, $S_4$, $S_5$, $S_6$, in this order. Different latin indices correspond to different replicas, and different greek letters for the same replica correspond to different colors. 
}
\label{potts_ampl}
\end{center} 
\end{figure}

Different theories are distinguished by their internal symmetries. The $q$-state Potts model \cite{Wu}, defined by the lattice Hamiltonian ${\cal H}=-\sum_{\langle i,j\rangle}J_{ij}\delta_{s_i,s_j}$, $s_i=1,2,\ldots,q$, is characterized by the $S_q$ symmetry corresponding to permutations of the $q$ values ("colors") that each site variable $s_i$ can take. For the pure ferromagnet ($J_{ij}=J>0$) below critical temperature the massive particle excitations of the field theory describing the scaling limit are kinks $A_{\alpha\beta}$ ($\alpha,\beta=1,2,\ldots,q$; $\alpha\neq\beta$) interpolating between pairs of degenerate ground states \cite{CZ}. The trajectory of $A_{\alpha\beta}$ in space-time corresponds to a domain wall separating a region with magnetization $\alpha$ from a region with magnetization $\beta$. At criticality the $q$ ferromagnetic phases coalesce and there are no kinks, but massless particles $A_{\alpha\beta}$ still provide the correct degrees of freedom \cite{paraf} and can be thought as yielding boundaries between clusters of spins with different colors. More generally, the particles $A_{\alpha\beta}$ can be shown to describe also  antiferromagnetic cases ($J<0$) \cite{DT}, and must be regarded as the basic way of representing $S_q$ symmetry in the scattering description. This symmetry leaves four inequivalent two-body amplitudes $S_0$, $S_1$, $S_2$ and $S_3$; they are shown in the upper part of figure~\ref{potts_ampl}, where the index $i$ must be ignored for the time being.

Quenched disorder is introduced regarding the couplings $J_{ij}$ as identical random variables and averaging with respect to them in the free energy $-\ln Z$ rather than in the partition function $Z$. As usual, writing $\ln Z$ as $\lim_{n\to 0}\frac{Z^n-1}{n}$ maps the problem onto the study of $n\to 0$ coupled replicas of the pure system. Within our scattering formalism this amounts to considering a theory with excitations $A_{\alpha_i\beta_i}$, where $i=1,2,\ldots,n$ labels the different replicas. Invariance under permutations of the replicas and of the colors within each replica yields the seven amplitudes $S_0,S_1,\ldots,S_6$ of figure~\ref{potts_ampl}; two-particle processes change the colors of at most two replicas, and these are the only ones we need to keep track of. Crossing symmetry relates the amplitudes which are exchanged under exchange of time and space directions as $S_0 =S_0^* \equiv  \rho_0$, $S_1 =S_2^*\equiv \rho\,e^{i\varphi}$, $S_3 =S_3^*\equiv \rho_3$, $S_4 =S_5^*\equiv \rho_4\,e^{i\theta}$, $S_6 =S_6^*\equiv \rho_6$, 
where we introduced a parameterization in terms of $\rho$ and $\rho_4$ non-negative, and $\rho_0,\rho_3,\rho_6,\varphi,\theta$ reals. The modulus square of an amplitude gives the probability that the given initial state scatters into the given final state. As a consequence the $S$-matrix (i.e. the matrix whose entries are the scattering amplitudes and are labeled by the initial and final states) is unitary, a property that in the present case results into the equations
\bea
&& \rho_3^2+(q-2)\rho^2+(n-1)(q-1)\rho_4^2=1\,,
\label{u1}\\
&& 2\rho\rho_3\cos\varphi+(q-3)\rho^2+(n-1)(q-1)\rho_4^2=0\,,
\label{u2}\\
&& 2\rho_3\rho_4\cos\theta+2(q-2)\rho\rho_4\cos(\varphi+\theta)+\nonumber\\
&& \hspace{.3cm}+(n-2)(q-1)\rho_4^2=0\,,
\label{u3}\\
&& \rho^2+(q-3)\rho_0^2=1\,,
\label{u4a}\\
&& 2\rho_0\rho\cos\varphi+(q-4)\rho_0^2=0\,,
\label{u4b}\\
&& \rho_4^2+\rho_6^2=1\,,
\label{u5}\\
&& \rho_4\rho_6\cos\theta=0\,.
\label{u6}
\eea
For example (\ref{u1}) follows from the fact that $1=\langle A_{\alpha_1\gamma_1}A_{\gamma_1\alpha_1}|\textrm{S}\textrm{S}^\dagger|A_{\alpha_1\gamma_1}A_{\gamma_1\alpha_1}\rangle=\sum_{j,\beta}|\langle A_{\alpha_1\gamma_1}A_{\gamma_1\alpha_1}|\textrm{S}|A_{\alpha_j\beta_j}A_{\beta_j\alpha_j}\rangle|^2$ yields $|S_3|^2$ for $j=1,\beta=\gamma$, a term $|S_2|^2$ for $j=1$ and each color $\beta\neq\alpha,\gamma$, and a term $|S_4|^2$ for each replica $j\neq 1$ and each color $\beta\neq\alpha$. 
Eqs.~(\ref{u1}-\ref{u6}) reduce to those of the pure model \cite{paraf} when $n=1$ and the equations which still contain $\rho_4$ and/or $\rho_6$ are ignored. Notice that $q$ and $n$ appear as parameters which can be given real values. Notice also that $\rho_4=0$ yields $n$ non-interacting replicas, since $S_4=S_5=0$ and, due to (\ref{u5}), $S_6=\pm 1$. We recall that in one spatial dimension scattering involves position exchange on the line, so that a scattering amplitude equal to 1 (resp. $-1$) corresponds to non-interacting bosons (resp. fermions). 

At a generic instant of time a two-particle excitation divides the line into a left, a central and a right region. We call neutral (resp. charged) the excitations for which the colors in the left and right regions are equal (resp. different). The neutral combination $\sum_{\gamma_i}A_{\alpha_i\gamma_i}A_{\gamma_i\alpha_i}$ scatters into itself with an amplitude
\EQ
S=S_3+(q-2)S_2+(n-1)(q-1)S_4\,,
\label{S}
\EN
which by unitarity is a phase. Similarly, the charged combinations $\sum_{\gamma_i}A_{\alpha_i\gamma_i}A_{\gamma_i\beta_i}$ and $A_{\alpha_i\beta_i}A_{\alpha_j\beta_j}+A_{\alpha_j\beta_j}A_{\alpha_i\beta_i}$ scatter into themselves with phases
$\Sigma=S_1+(q-3)S_0$ and $\sigma=S_5+S_6$, respectively.

Passing to solutions, consider first the Ising case. With two colors available the amplitudes $S_0,S_1,S_2$ are unphysical, and the unitarity equations still containing $\rho_0$ and/or $\rho$ after setting $q=2$ can be ignored. The remaining equations give the solutions
\bea
&& \rho_3=-1\,,\hspace{.4cm}\rho_4=0\,,\hspace{.4cm}\rho_6=\pm 1 \,,
\label{ising_decoupled}\\
&& \rho_3=2\cos\theta=-\sqrt{2-n}\,,\hspace{.4cm}\rho_4=1\,,\hspace{.4cm}\rho_6= 0 \,,
\label{ising_coupled}
\eea
where we took into account that in 2D the pure Ising model is a free fermionic theory, so that we must have $S_3=-1$ for $n=1$; this eliminates the doubling related to the fact that, given a solution of the crossing and unitarity equations, another solution is obtained reversing the sign of all amplitudes. The solution with $\rho_4=0$ corresponds, for any real $n$, to $n$ decoupled Ising FPs. The solution with $\rho_4=1$ is defined for $-2\leq n\leq 2$ and corresponds to strongly coupled replicas; for $n\to 0$ it can account for a strong disorder FP such as the Nishimori point \cite{Nishimori}. 

Back to $q$ generic, we restrict to coupled replicas, i.e. $\rho_4\neq 0$, and first consider the case in which (\ref{u6}) is solved taking $\cos\theta=0$. Then (\ref{u3}) shows that $\rho_4\to 0$ as $q\to 2$, namely for $q\to 2$ the decoupled solution (\ref{ising_decoupled}) and Ising critical exponents are obtained. We are then considering, within an exact framework, the case studied perturbatively for $q\to 2$ in \cite{Ludwig,DPP}. There exists and is unique a solution defined for all $q\geq 2$, which then corresponds to the line of critical points studied numerically in \cite{CJ} in the range $2\leq q\leq 8$. More precisely, this solution is defined for $q\geq\sqrt{2}$, and for $n=0$ reads
\bea
&& \cos\theta=\rho_0=0\,,\hspace{.6cm}\rho=1\,,\hspace{.6cm}\rho_3=2\cos\varphi=-\frac{2}{q}\,,\nonumber\\
&&\rho_4^2=1-\rho_6^2=\frac{(q-2)^2(q+1)}{q^2(q-1)}\,.
\label{III2a}
\eea
Notice that $\rho_4$ monotonically increases from 0 at $q=2$ to 1 as $q\to\infty$, so that the solution interpolates from weak to strong disorder.  
It superficially seems to correspond to an ordinary $q$-dependent line of FPs, but it follows from (\ref{S}) and (\ref{u3}) that for $\cos\theta=n=0$ the imaginary part of $S$ vanishes, so that $S=-1$ for all $q$; on the other hand, the phases associated to the charged channels have $\textrm{Im}\,\Sigma=\rho\sin\varphi$ and $\textrm{Im}\,\sigma=\pm \rho_4$, and are $q$-dependent.  This means that, remarkably, along the critical line (\ref{III2a}) a specific symmetry sector of the theory, the one to which the combination $\sum_{\gamma_i}A_{\alpha_i\gamma_i}A_{\gamma_i\alpha_i}$ belongs, becomes $q$-independent {\it at $n=0$}, i.e. only in the limit required for quenched disorder. 

To understand the consequences for critical exponents we need to consider the symmetry properties of the operators in the $q$-state Potts model. The spin operator has components $\sigma_{\beta}(x)=\delta_{s(x),\beta}-1/q$, $\beta=1,2,\ldots,q$; what we say separately applies to each replica and we omit the replica index in order to simplify the notation. The energy density operator $\varepsilon(x)$ is the most relevant operator appearing in the operator product expansion $\sigma_\beta\cdot\sigma_\beta$, and is $S_q$-invariant. Both $\varepsilon$ and $\sigma_\beta$, when acting on the vacuum of the field theory, create neutral excitations; indeed charged excitations, which interpolate between different colors, are non-local with respect to the spin operator and are created by disorder-like operators. In this respect, the actual difference between $\varepsilon$ and $\sigma_\beta$ is the following: $\varepsilon$ is $S_q$-invariant and creates two-particle excitations $A_{\alpha\gamma}A_{\gamma\alpha}$ with the same, $\alpha$- and $\gamma$-independent coefficient; $\sigma_\beta$ creates $A_{\alpha\gamma}A_{\gamma\alpha}$ with different coefficients for $\beta=\alpha$, $\beta=\gamma$, and $\beta\neq\alpha,\gamma$. In the pure model this difference makes more difficult the determination of off-critical spin correlations \cite{DVC} with respect to that of energy correlations \cite{DC}. For our present purposes it implies that $\varepsilon$ creates the combination $\sum_\gamma A_{\alpha\gamma}A_{\gamma\alpha}$ as a whole and, as any other $S_q$-invariant neutral operator, belongs to the sector of the theory which is $q$-independent along the critical line (\ref{III2a}). The scaling dimensions of these operators keep along the line the Ising value they have at $q=2$; in particular, the exponent $\nu$, which is determined by the dimension of $\varepsilon$, keeps the value 1 along the line. Conversely, the operators which are not $S_q$-invariant have dimensions which change along the line; in particular, the exponent $\beta$, which is determined by the dimensions of $\sigma_\alpha$ and $\varepsilon$, is $q$-dependent. 

These findings indicate that the numerical \cite{CFL,DW} and theoretical \cite{KSSD} studies pointing at a $q$-independent exponent $\nu$ with Ising value, and the theoretical \cite{Ludwig,DPP} and numerical \cite{CJ,CB} results yielding a $q$-dependent exponent $\beta$ do actually correspond to the {\it same} critical line. The weak variation of $\nu$ found in \cite{CJ} always has the Ising value 1 within error bars, and  the perturbative expansion of \cite{Ludwig,DPP} yields $\nu\approx 1.02$ if evaluated at $q=3$, a result still very close to the Ising value. The reason why this latter result is reliable is that the actual expansion parameter is the deviation of $\nu$ from the Ising value in the pure models, and this is still small at $q=3$. The suggestion of \cite{KSSD,CJ} that the strong randomness, large $q$ analysis of \cite{KSSD} and the weak randomness, $q\to 2$ analysis of \cite{Ludwig,DPP} correspond to different critical lines turns out to be unnecessary: the same critical line (\ref{III2a}) can account for constant $\nu$ and varying $\beta$, and interpolates from weak to strong randomness as $q-2$ grows from small to large values. The value $\nu=1$ proposed in \cite{AdI} at $q=\infty$ also agrees with our result. At first sight the constance of $\nu$ appears incompatible with the $q$-dependent effective central charge $c^\prime$ found numerically in \cite{Picco,CJ}. Indeed, the central charge $c$ is related to the stress-energy tensor, and then to the $S_q$-invariant sector responsible for the $q$-independence of $\nu$. However, for systems with quenched disorder the quantity measured from the finite size dependence of the free energy \cite{BCN,Affleck} is the {\it effective} central charge $c^\prime=\partial_nc(n)|_{n=0}$. Since the $S_q$-invariant sector of the theory becomes $q$-independent only at $n=0$, $c^\prime$ is $q$-dependent, in agreement with the numerical results. The central charge itself is $\lim_{n\to 0}n/2=0$ at the decoupling point $q=2$, and keeps this value along the line. 

When (\ref{u6}) is solved taking $\rho_6=0$, (\ref{u5}) implies $\rho_4=1$ and strong coupling solutions are obtained which reduce to (\ref{ising_coupled}) for $q=2$. Among these, only two are defined for all $q\geq 2$ (and actually for all $q$), and for $n=0$ have
\EQ
\rho_0=\rho_6=0,\hspace{.4cm}\rho=\rho_4=1,\hspace{.4cm}\rho_3=2\cos\varphi=-\sqrt{2},
\label{IIa}
\EN
and $\theta$ determined by (\ref{u3}); one of these solutions has $\cos\theta=\cos\varphi$ and is completely $q$-independent.  All solutions of the equations (\ref{u1})-(\ref{u6}) will be listed in \cite{DT}. It is worth stressing that the inputs of our formalism are conformal invariance and internal symmetry, $S_q$ in the present case, so that the space of solutions of the crossing and unitarity equations contains the ferromagnetic case as well as (lattice-dependent) antiferromagnetic and mixed realizations. 

It was shown in \cite{paraf} for the pure model how the scattering formalism essentially contributes to a self-contained field theoretical determination of the exponents, in a context in which sufficient insight is available about the underlying CFT. On the other hand, the results of this paper say that the CFTs of quenched random criticality may admit symmetry-independent sectors and are of a peculiar type; those we are providing are the first exact properties. In perspective, it should be possible to gain further insight on the conformal properties and, through the combination with the scattering solution, to determine the non-Ising exponents along the lines of \cite{paraf}. 

Since the formalism is generally applicable in 2D, we consider as a further illustration the case of the XY model. The corresponding symmetry $O(2)\sim U(1)$ is represented by a pair of particles $A$ and $\bar{A}$ with $U(1)$ charge 1 and $-1$, respectively. Writing the amplitudes for the scattering processes allowed by charge conservation and considering $n$ replicas one obtains a system of unitarity equations which coincides with (\ref{u1})-(\ref{u6}) with $q=3$ and (\ref{u4b}) omitted ($S_0$ is unphyscal for $q=3$). The reason for this coincidence is that the permutational group $S_3$ amounts to $Z_3$ cyclic permutations times a $Z_2$ reflection. As a consequence the Potts excitations $A_{\alpha\beta}$ admit the identifications $A_{\alpha,\alpha+1(\textrm{mod}\,3)}\equiv A$, $A_{\alpha,\alpha-1(\textrm{mod}\,3)}\equiv\bar{A}$, which map the three-state Potts amplitudes onto the $U(1)$ amplitudes; crossing and unitarity then yield the same equations. It is worth stressing that this coincidence does not extend to critical exponents. The XY spin operator is $U(1)$-charged and creates charged excitations; it clearly differs from the Potts spin operator $\sigma_{\alpha}$. These operators, as well as the energy operators, are discussed in relation with the scattering formalism in \cite{paraf,fpu} for the pure models. 

Having understood that we can refer to the Potts equations with $q=3$, consider first the pure case, i.e. $n=1$. Since (\ref{u4a}) fixes $\rho=1$, (\ref{u1}) gives $\rho_3=0$. Then (\ref{u2}), which reads $\rho_3\cos\varphi=0$, is identically satisfied. Since (\ref{u3}) plays no role in the pure model, we see that $\varphi$ remains as a free parameter labeling a critical line in the $U(1)$ model. This is the line of FPs onto which the pure XY ferromagnet renormalizes below the Kosterlitz-Thouless \cite{KT} transition temperature $T_{KT}$. Scale invariance below $T_{KT}$ forbids a spontaneous magnetization and leads to power-law decay of correlations ("quasi-long-range order"), consistently with the absence of spontaneous breaking of continuous symmetries in two dimensions \cite{MW,H}. 
When passing to interacting replicas ($\rho_4\neq 0$), (\ref{u1}) requires $n<1$. The solutions are the $q=3$ case of those discussed for the Potts model, and do not admit free parameters, meaning that the only critical line is that corresponding to non-interacting replicas. Since the pure model is already unable to order, the FPs that the equations yield at $n=0$ are not expected to provide a positive temperature transition point for the random bond XY ferromagnet. These solutions, however, can account for elimination of  quasi-long-range order by sufficiently strong  bond disorder, as found  within a strong coupling approximation in \cite{Jose}.

In summary, we have shown how properties of 2D systems with quenched bond disorder can be studied exactly, directly in the continuum limit and at criticality, within the framework of scale invariant scattering theory. For the $q$-state Potts model the analysis shows that there is a line of FPs along which the disorder strength vanishes at $q=2$ and then increases continuously with $q$, and reveals a mechanism allowing the exponent $\nu$ and the central charge to stay constant while $\beta$ and the effective central charge vary. These unusual features of the theory account for numerical and theoretical results that had not seemed all compatible with each other.




\begin{thebibliography}{99}
\bibitem{Harris} A.B. Harris, J. Phys. C 7 (1974) 1671.
\bibitem{HB} K. Hui and A.N. Berker, Phys. Rev. Lett. 62 (1989) 2507. 
\bibitem{AW} M. Aizenman and J. Wehr, Phys. Rev. Lett. 62 (1989) 2503.
\bibitem{FK} P.W. Kasteleyn and E.M. Fortuin, J. Phys. Soc. Japan Suppl. (1969) 2611; Physica 57 (1972) 536.
\bibitem{Baxter} R.J. Baxter, Exactly Solved Models of Statistical Mechanics, Academic Press, London, 1982.
\bibitem{DD} V.S. Dotsenko and Vl. S. Dotsenko, Sov. Phys. JETP Lett. 33 (1981) 37; Adv. Phys. 32 (1983) 129.
\bibitem{Ludwig} A.W.W. Ludwig, Nucl. Phys. B 330 (1990) 639.
\bibitem{DPP} V. Dotsenko, M. Picco and P. Pujol, Nucl. Phys. B 455 (1995) 701. 
\bibitem{CFL} S. Chen, A.M. Ferrenberg and D.P. Landau, Phys. Rev. Lett. 69 (1992) 1213; Phys. Rev. E 52 (1995) 1377. 
\bibitem{DW} E. Domany and S. Wiseman, Phys. Rev. E 51 (1995) 3074.
\bibitem{KSSD} M. Kardar, A.L. Stella, G. Sartoni and B. Derrida, Phys. Rev. E 52 (1995) R1269.
\bibitem{CJ} J. Cardy and J.L. Jacobsen, Phys. Rev. Lett. 79 (1997) 4063.

J.L. Jacobsen and J. Cardy, Nucl. Phys. B 515 (1998) 701.
\bibitem{Picco} M. Picco, Phys. Rev. Lett. 79 (1997) 2998.
\bibitem{JP} J.L. Jacobsen and M. Picco, Phys. Rev. E 61 (2000) R13. 
\bibitem{AdI} J.-Ch. Angl\`es d'Auriac and F. Igl\'oi, Phys. Rev. Lett. 90 (2003) 190601.
\bibitem{paraf} G. Delfino, Annals of Physics 333 (2013) 1.
\bibitem{ELOP} R.J. Eden, P.V. Landshoff, D.I. Olive and J.C. Polkinghorne, The analytic S-matrix, Cambridge, 1966.
\bibitem{fpu} G. Delfino, Annals of Physics 360 (2015) 477.
\bibitem{Wu} F.Y. Wu, Rev. Mod. Phys. 54 (1982) 235.
\bibitem{CZ} L. Chim and A.B. Zamolodchikov, Int. J. Mod. Phys. A 7 (1992) 5317.
\bibitem{DT} G. Delfino and E. Tartaglia, in preparation.
\bibitem{Nishimori} H. Nishimori, J. Phys. C 13 (1980) 4071; Prog. Theor. Phys. 66 (1981) 1169.
\bibitem{DVC} G. Delfino, J. Viti and J. Cardy,  J. Phys. A 43 (2010) 152001.
\bibitem{DC} G. Delfino and J. Cardy, Nucl. Phys. B 519 (1998) 551.
\bibitem{CB} C. Chatelain and B. Berche, Phys. Rev. Lett. 80 (1998) 1670.
\bibitem{BCN} H.W. Blote, J.L. Cardy, and M.P. Nightingale, Phys. Rev. Lett. 56 (1986) 742.
\bibitem{Affleck} I. Affleck, Phys. Rev. Lett. 56  (1986) 746.
\bibitem{KT} J.M. Kosterlitz, D.J. Thouless, J. Phys. C: Solid State Phys. 6 (1973) 1181.
\bibitem{MW} N.D. Mermin, H. Wagner, Phys. Rev. Lett. 17 (1966) 1133.
\bibitem{H} P.C. Hohenberg, Phys. Rev. 158 (1967) 383.
\bibitem{Jose} J.V. Jos\'e, Phys. Rev. B 20 (1979) 2167.
\end{thebibliography}
\end{document}